\documentclass[preprint2]{aastex}
\input epsf.sty

\def\Referencja#1{\noindent\hangindent=50pt\hangafter=1\nobreak
\frenchspacing #1 \par}

\shorttitle{Intermittent activity in AGN}
\shortauthors{Janiuk et al.}

\begin{document}

\title{On the turbulent $\alpha$-disks and the intermittent activity in AGN}
\author{Agnieszka Janiuk and Bo\.zena Czerny}
\affil{ Nicolaus Copernicus Astronomical Center, Bartycka 18,
            00-716 Warsaw, Poland}
\email{agnes@camk.edu.pl, bcz@camk.edu.pl} 

\author{Aneta Siemiginowska}
\affil{Harvard-Smithsonian Center for Astrophysics, 60 Garden
Street, MA02138, Cambridge, USA}
\email{asiemiginowska@cfa.harvard.edu}

\author{Ryszard Szczerba}
\affil{Nicolaus Copernicus Astronomical Center, Rabia\'nska 8,
            87-100 Toru\'n, Poland}
\email{szczerba@ncac.torun.pl} 

%\clearpage

\begin{abstract}

We consider effects of the MHD turbulence on the viscosity during the
evolution of the thermal-viscous ionization instability in the
standard $\alpha$-accretion disks. 
We consider the possibility that the accretion onto a supermassive
black hole proceeds through an outer standard accretion disk and
inner, radiatively inefficient and advection dominated flow.  In this
scenario we follow the time evolution of the accretion disk in which
the viscosity parameter $\alpha$ is constant throughout the whole
instability cycle, as implied by the strength of MHD turbulence.  We
conclude that the hydrogen ionization instability is a promising
mechanism to explain the intermittent activity in AGN.

\end{abstract}


\section{Introduction}

The standard accretion disk (Shakura \& Sunyaev 1973) is known to be
subject to the thermal - viscous instability due to the partial
hydrogen ionization (Meyer \& Meyer-Hofmeister 1981; Smak 1984). As a
result of this instability the disk cycles between the two states: a
hot and mostly ionized state with a large local accretion rate and a
cold, neutral state with a low accretion rate. This instability has
originally been proposed to explain the large amplitude luminosity
variations observed in cataclysmic variables (CV; Smak 1982; Meyer \&
Meyer-Hofmeister 1982).  It is also believed that the same mechanism
is responsible for the eruptions in soft X-ray transients (SXT; see
e.g. Cannizzo, Ghosh \& Wheeler 1982; Dubus, Hameury \& Lasota 2001;
Lasota 2001 for review). The ionization instability was also shown to
operate in the disk around a supermassive black hole in active 
galactic nuclei (AGN; Lin \& Shields
1986; Clarke 1988; Mineshige \& Shields 1990; Siemiginowska, Czerny \&
Kostyunin 1996).

The characteristic timescales of a cycle activity scale roughly with
the mass of a compact object (Hatziminaoglou et al. 2001). Therefore,
the observed timescale of cycle of order of years in binaries
translates into thousands to millions of years in galaxies, which
harbor a supermassive black hole.

The variability amplitudes and timescales observed in Galactic binary
systems require the viscosity to be smaller in the quiescent disk than
during the outburst.  Various {\it ad-hoc} viscosity scaling laws have
been used in the disk evolution models, including the simplest one
with a constant viscosity, however usually $\alpha$ is assumed to be
4-5 times larger during the outburst than in the quiescence (Cannizzo
1993). The value of $\alpha_{\rm hot}$ determines the timescale of an
outburst, while $\alpha_{\rm cold}$ governs the separation between the
subsequent outbursts. The difference between these two parameters
determines the outburst amplitude.

The dependence of the viscosity parameter on the disk state is
suggested to come out as a property of the magneto-rotational
instability (MRI) mechanism (Balbus \& Hawley 1991; 1998). Nonlinear
development of this instability, which is assumed to be a primary
source of the disk turbulence, is sensitive to the presence of
resistive diffusion of magnetic field (Hawley, Gammie \& Balbus 1996),
measured by the magnetic Reynolds number $Re_{\rm M}$. Gammie \& Menou
(1998) calculated the Reynolds number for the two states of the disk
in CVs and show that it is low in the
quiescence. It means that the magneto-hydrodynamical (MHD) turbulence
dies away and the matter accumulates in the outer disk.

Menou \& Quataert (2001) argued that the Reynolds number will not be
low enough to suppress the MHD turbulence in the quiescent disk of AGN.
Therefore the efficiency of angular momentum transport should be
comparable in the hot and cold states ($\alpha_{\rm cold} \approx
\alpha_{\rm hot}$).  In such a case the outburst amplitude is
dramatically reduced, and the ionization instability leads only to a
small amplitude flickering (Siemiginowska et al. 1996).

The conclusion about the small amplitude of outbursts was obtained,
however, for a geometrically thin, optically thick disk extending down
to the marginally stable orbit. This assumption may not be correct.
When the local accretion rate is low, the local radiative cooling may
not be efficient (Rees et al. 1982; Begelman 1985). The innermost part
of the disk may be replaced with an optically thin, hot and possibly
two-temperature plasma. An example of such a solution, advection dominated
accretion flow (ADAF) was introduced by Ichimaru (1977) and Narayan \&
Yi (1994). The development of the inner flow through the disk evaporation
was recently discussed by several authors, starting with Meyer \&
Meyer-Hofmeister (1994).

In this paper we  first perform a self-consistency check of the
$\alpha$ description of the angular momentum transfer. Having justified
$\alpha_{\rm cold}$ = $\alpha_{\rm hot}$ approach to AGN accretion disks,
we consider the time evolution of the disk under the influence of the 
ionization instability. We also take into account the evaporation of the 
inner disk.

We define a parameter space where
the instability zone (i.e. the partial hydrogen ionization zone) and the
evaporation region overlap.
The overlapping of these two regions strongly enhances the outburst
amplitude due to the time evolution of the disk evaporation radius.
We illustrate this behavior with exemplary lightcurves.

The structure of this article is as follows. In Section
\ref{sec:reynolds} we define the criterion for efficient development
of the MHD turbulence in the accretion disk, taking into account its
vertical structure. We confirm that indeed the bi-modal viscosity
behavior is characteristic for Galactic binary systems. However, a
single value of the viscosity parameter is appropriate in the case of
AGN disks.  In Section
\ref{sec:evap} we check if the zone of partial hydrogen ionization
overlaps with the disk evaporation region and we calculate the
evolution of the disk luminosity in case of the evaporated inner disk.
Finally, in Section \ref{sec:diss} we discuss our results and give
conclusions.

\section{Magneto-rotational turbulence}
\label{sec:reynolds}

In this Section we estimate the strength of the MHD turbulence in the
$\alpha$ accretion disks and investigate whether the viscosity
prescription with a constant parameter $\alpha$ is appropriate for
both Galactic systems (CVs and SXTs) and AGN.  We only
consider the instability due to partially ionized hydrogen and do not
include regions affected by the radiation pressure instability (see
Janiuk, Czerny \& Siemiginowska 2002 for that model details).  We assume
that the viscous stress tensor is proportional to the gas pressure
$P_{\rm gas}$ and therefore there is no radiation pressure dominated
branch on the stability figures (so-called S-curves) described below.

We use the dimensionless accretion rate in Eddington units, 
$\dot m = \dot M / \dot M_{\rm Edd}$, assuming the
efficiency of 1/12, as implied for the Schwarzschild  black hole
by the Newtonian potential:
\begin{equation}
\dot M_{\rm Edd} = 3.52  { M \over 10^8 M_{\odot}} [M_{\odot}/yr].
\end{equation}

The disk vertical structure (indicated by $z$) is calculated by
solving the equations of viscous energy dissipation, hydrostatic
equilibrium, and energy transfer:
\begin{equation}
{d F \over d z} = \alpha P_{\rm gas} (-{d\Omega \over d r})
\end{equation}
\begin{equation}
{1 \over \rho} {d P \over d z} = -\Omega^{2} z
\label{eq:hydro}
\end{equation}
\begin{equation}
{d T \over d z} = -{3 \kappa \rho \over 4 a c T^{3}} F_{\rm l}
\end{equation}
where we solve for temperature, density and pressure profiles, $T(z)$, $\rho(z)$ and $P(z)$.
Here $\Omega$ is the Keplerian angular velocity, $a$ and $c$ are physical constants.
$F_{\rm l}$  
is the energy flux transported locally in the direction perpendicular 
to the equatorial plane and carried either by radiation or by convection:
\begin{equation}
F_{\rm l} = F_{\rm rad} ~~~ \nabla_{\rm rad} \le \nabla_{\rm ad},  \\
~~~~~~ F_{\rm l} = F_{\rm rad} +F_{\rm conv} ~~~ \nabla_{\rm rad} > \nabla_{\rm ad}
\end{equation}
The frequency-averaged opacity $\kappa$ is taken to
be the Rosseland mean and includes the electron scattering, free-free
and bound-free transitions. The opacity tables are from Alexander,
Johnson \& Rypma (1983) and Seaton et al. (1994).  The presence of
dust and molecules is included in the opacity description.  The
details of the model were discussed in Pojma\'nski (1986) and 
R\'o\.za\'nska et al. (1999).

\begin{figure}
\epsfxsize = 80 mm 
\epsfbox[40 150 600 700]{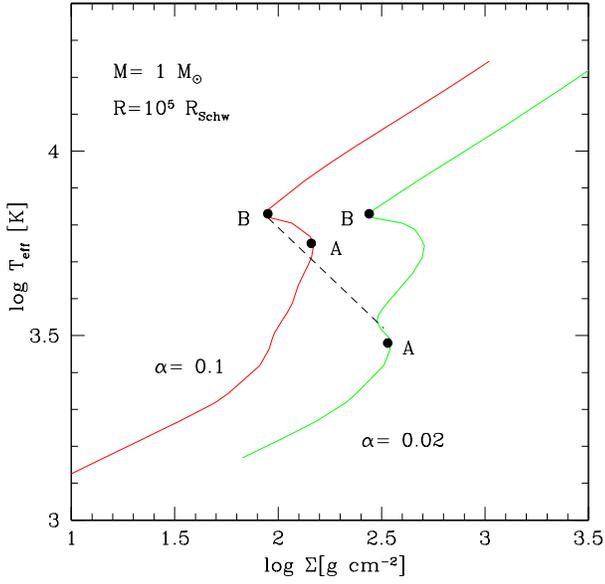}
%\plotone{f1.eps}
\caption{The stability curves calculated for the disk around $M= 1 M_{\odot}$ 
at the radius $R=10^{5} R_{\rm Schw}$.  The viscosity
parameters are $\alpha = 0.1$ (left curve) and $\alpha = 0.02$ (right
curve). The dashed line marks the combined stability curve,
which would result from the
assumption that $\alpha_{\rm cold} = 0.02$ (on the lower stable branch) 
and $\alpha_{\rm hot} = 0.1$ (on the upper stable branch).  
Point $A$ indicates the lowest location of the unstable disk
and a starting point of the ionization instability cycle 
and the instability ends in the point $B$.}
\label{fig:s_CV}
\end{figure}

\begin{figure}
\epsfxsize = 80 mm 
\epsfbox[40 150 600 560]{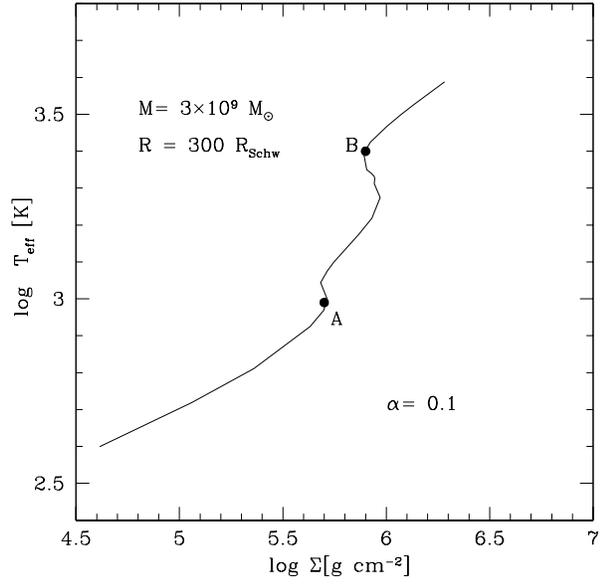}
%\plotone{f2.eps}
\caption{The stability 
curve calculated for the disk around a supermassive black hole of $M=3
\times 10^{9} M_{\odot}$ at the radius $R=300 R_{\rm Schw}$. The
viscosity parameter is $\alpha = 0.1$.  The point $A$ is the starting
point of the ionization instability cycle
and the instability ends in the point $B$.
\label{fig:scurve}}
\end{figure}

Figures \ref{fig:s_CV} and \ref{fig:scurve} show, in $\Sigma$
vs. $T_{\rm eff}$ (surface density vs. effective temperature) plane,
the local solutions of the disk vertical structure calculated for a
range of accretion rates  for two extreme cases of the mass
of the central object: 1M$_{\odot}$ and 3 $\times$
10$^9$M$_{\odot}$. The solutions located on both lower (below point A) 
and upper (above point B) 
branches of the S-curves are stable.  The slope of the middle branch is,
at least in some parts, negative which means that at that range of accretion rates
the disk is thermally and viscously unstable.  

Note, that in case of CVs (Figure \ref{fig:s_CV}) the location of the starting point 
of the instability is very sensitive to the adopted value of $\alpha$. For a small
viscosity parameter, $\alpha=0.02$, the point $A$ is at much lower temperature than
in case of large $\alpha=0.1$. The latter 'smoothens' the S-curve in its lower part,
so that only a single 'wiggle' remains. Therefore the point $A$, defined as the first
critical point above which the S-curve slope becomes negative, is shifted upwards in 
temperature in comparison with that characteristic for small viscosities.
This fact should be taken into account when modeling
the limit cycle in dwarf novae by means of the 'combined' S-curves with 
$\alpha_{cold} < \alpha_{hot}$.

This behavior is however characteristic only for accretion disks around 1$M_{\odot}$,
while vanishes already for SXTs. Also for supermassive black holes there are always
two 'wiggles' in the S-curve, regardless of the viscosity parameter.

%Point $A$ in both figures indicates the parameters
%where the disk becomes unstable.  The choice of rather large radius
%for a low mass object and a small radius for a supermassive black hole
%was motivated by the radial location of the instability zone.

 The ionization instability is characteristic for all accretion disks but the range
of accretion rates (i.e. the location
of the S-curve on the $T-\Sigma$ plane) 
depends on the chosen disk radius. We can invert this problem
and say that for a fixed external accretion rates the instability will appear
only for a certain range of radii. At larger radii the disk is on the lower 
stable branch while at lower radii it is on the higher, also stable branch.
Simple analytical formulae for the unstable zone were provided by Siemiginowska
et al. (1996) and applied by Menou \& Quataert (2001) in their analysis.

In Figure \ref{fig:topo} we show the radial extension of the unstable 
zone as a function of the accretion rate (in Eddington units)  
 calculated numerically from our disk model. The position of the
instability zone depends on the mass of the central object so we choose
values representative for all objects from CVs to AGN with a large black hole mass.
%for the case of a supermassive black hole ($M=10^{6} M_{\odot}$
%and $M=10^{8} M_{\odot}$).  
The plot shows the radial extension as a
function of the accretion rate (in Eddington units).  The horizontal
slices of the shaded region correspond to the accretion rates, for
which the disk is unstable at a particular radius.
%AGN disks are in quiescence for $\dot m < 10^{-3}$. Therefore in the subsequent calculations 
%we choose the location of order of $300 R_{Schw}$.
%Note, that in case of a stellar mass black hole system the whole unstable zone will
%be shifted outwards in radius, since the disk temperature scales as $M^{-1/4}$.
For $\dot m \approx 0.01$ (an accretion rate typical for many objects)
the instability zone is located at $\sim 10^{5} R_{\rm Schw}$ for a CV,
at $\sim 2
\times 10^{3} R_{\rm Schw}$ for $M \sim 10^{6} M_{\odot}$ and at $\sim
500 R_{\rm Schw}$ for an extremely massive black hole of $M= 3 \times
10^{9} M_{\odot}$. 
%Therefore the S-curve shown in Figure
%\ref{fig:s_CV} is indeed representative for a typical behavior of the CV 
%systems, while Figure \ref{fig:scurve} is a typical for a very massive
%AGN with a relatively low mean accretion rate $\dot m \approx
%10^{-3}$.

Time evolution of the unstable part of the disk proceeds roughly in the
form of oscillations between the upper and the lower stable branches. As argued
by Gammie \& Menou (1998), high temperature upper branch conditions are always
favorable for the development of the efficient MRI and high values
of viscosity are appropriate there. Lower branch conditions are different and
the MRI mechanism may not be efficient. Therefore, we perfom our self-consistency
check paying particular attention to the lower branch solutions at the vicinity
of the turning point $A$.

\subsection{Coupling of the magnetic field with the gas}

The behavior of the magnetic field is governed by the fluid
conductivity $\sigma$.  The time-dependent magnetic field in the disk
is described by the equation (Parker 1979):
\begin{equation}
{\partial B \over \partial t} = \nabla \times (\vec v \times \vec B) 
+ \eta \nabla^{2} \vec B
\label{eq:magndif}
\end{equation}
where $\eta \equiv c^{2}/4\pi\sigma$ denotes resistivity.  The
characteristic diffusion time in which the initial configuration of
the magnetic field will decay is equal to $\tau = L^{2}/\eta$, where $L$
indicates the characteristic spatial scale. For timescales short in
comparison with $\tau$ the second term in Equation
\ref{eq:magndif} can be neglected and the magnetic field lines are frozen
into the gas. Magnetic Reynolds number, defined as:
\begin{equation}
Re_{\rm M}={v \tau \over L}
\label{eq:magn}
\end{equation}
can be used to distinguish between two cases: (1) the diffusion of
field lines within the disk  and (2) the lines are frozen in and
carried along with the matter.  Here we identify the velocity $v$ with
the sound speed in the disk and the length $L$ of the magnetic field
spatial variations with the disk thickness, as commonly used in the
simulations (see e.g. Hawley et al. 1996; Gammie \& Menou 1998).

Apart from the ohmic diffusion also the ambipolar diffusion may be
important. The ambipolar Reynolds number is defined as:
\begin{equation}
Re_{\rm A}={\nu_{\rm ni} \over \Omega}
\label{eq:amb}
\end{equation}
where $\nu_{\rm ni}$ is the frequency of collisions between ions and
neutral particles.

We can calculate the both Reynolds numbers locally in the disk. 
In order to determine the resistivity, $\eta = (c^{2} m_{\rm e} / 4
\pi e^{2}) \times (\nu_{\rm en} / n_{\rm e})$, we have to estimate the
number density of electrons $n_{\rm e}$ and neutral particles $n_{\rm
n}$. The frequency of collisions between electron and neutral
particles is defined as $\nu_{\rm en} = 8.3
\times 10^{-10} T^{1/2} n_{\rm n}$ in s$^{-1}$ (Draine, Roberge \&
Dalgarno 1983).  The ionization
fraction $x_{\rm e} = n_{\rm e}/n_{\rm n}$ is calculated from the Saha
equation for any given value of temperature and density in the
disk. In this temperature range the free electrons originate mostly
from the ionization of hydrogen, sodium, potassium and calcium.

The magneto-rotational instability acts over the entire disk
thickness. Therefore, we need to consider the entire vertical disk
structure when we estimate the coupling of the magnetic field to the
gas.
This is why instead of calculating $Re_{\rm A}$ and $Re_{\rm M}$ at
the equatorial plane, we introduce the density weighted values of
these parameters, integrated over the disk height:
\begin{equation}
\bar Re_{\rm A} = {\int_{0}^{H} Re_{\rm A} \rho(z) dz \over 
\int_{0}^{H} \rho dz}
\end{equation}
and
\begin{equation}
\bar Re_{\rm M} = {\int_{0}^{H} Re_{\rm M} \rho(z) dz \over
\int_{0}^{H} \rho dz}
\end{equation}
where $\rho (z)$ is the density profile determined from the vertical
structure model and $H \equiv z_{\rm max}$ is the disk thickness.

Numerical simulations (Hawley et al. 1996; Hawley \& Stone 1998) show
that below the critical Reynolds numbers (defined as in Equations \ref{eq:magn} and \ref{eq:amb}):
\begin{equation}
Re_{\rm M}^{\rm crit} = 10^{4}
\end{equation}
and
\begin{equation}
Re_{\rm A}^{\rm crit} = 100
\end{equation}
the action of MHD turbulence is inefficient and cannot be important in
the angular momentum transport in the disk.  As a consequence, the
viscosity parameter ($\alpha \approx \alpha_{\rm mag}$) is very small.

If the disk Reynolds numbers drop below critical values in the
quiescence (e.g.  on the lower branch of the S-curve), while rise
above the critical values in the outburst (e.g. on the upper branch of
the S-curve) then the assumption that $\alpha_{\rm cold}
\ll \alpha_{\rm hot}$ (see Figure \ref{fig:s_CV}) is justified. 
Otherwise, the viscosity parameter should be the same on both cold and
hot branches of the S-curve (see Figure
\ref{fig:scurve}). Below we calculate the Reynolds numbers in the
quiescent disk and test them against the critical values in different
types of accreting systems.

\subsection{Results}

\subsubsection{Accretion disks in binary systems}

\begin{figure}
\epsfxsize = 80 mm 
\epsfbox[60 160 560 500]{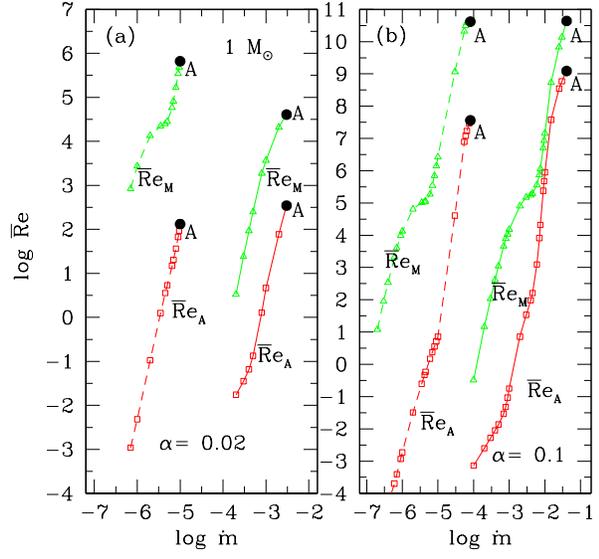}
%\plotone{f3.eps}
\caption{The density weighted magnetic and ambipolar Reynolds numbers 
calculated for the central mass $M=1 M_{\odot}$, plotted for two disk radii $R=10^{4}
R_{\rm Schw}$ (dashed line) and $R=10^{5}
R_{\rm Schw}$ (solid line),  as a function of
the accretion rate in the range corresponding to the cold branch of
the S-curve (cf. Fig. \ref{fig:s_CV}). The viscosity is $\alpha = 0.02$ (panel a) 
and $\alpha = 0.1$ (panel b).
\label{fig:re_CV}}
\end{figure}

We consider an accretion disk in CVs, e.g. around a white dwarf with a
mass M=1M$_{\odot}$ and calculate the disk vertical structure at 
two exemplary radii: 10$^4$R$_{\rm Schw}$ and   
10$^5$R$_{\rm Schw}$. Figure \ref{fig:re_CV} (a) 
shows the magnetic and
ambipolar Reynolds numbers for the case of small viscosity parameter 
$\alpha = 0.02$. It is plotted as a function of the accretion rate
corresponding to the lower branch of the S-curve (e.g. in Figure
\ref{fig:s_CV} below the point $A$).  Point $A$ indicates the highest
accretion rate on the lower, stable branch of the S-curve. Both
Reynolds numbers are  below the critical
values for most part of the lower branch. Magnetic Reynolds number is
somewhat above the critical value close to the point $A$.  The ambipolar
Reynolds number is well below the critical value at the smallest 
radius and approaches it at the point A only for the largest radius. 
In general, the magnetic number at the lower branch is almost independent
from the radius. The ambipolar number is increasing with the radius. However, 
the value of 10$^5 $R$_{\rm Schw}$ is rather an upper limit for the disk
size in CV systems so any values of the Reynolds number larger than those
in Figure \ref{fig:re_CV} are not expected.
 
In Figure \ref{fig:re_CV}  (b)  we plot the magnetic and
ambipolar Reynolds numbers up to the point $A$ for large value of $\alpha = 0.1$.
In this case the Reynolds numbers are mostly small,
but as the point $A$ reaches this time higher temperatures, also their values
can exceed critical ones in the end of the extended stable branch. 

Therefore, whenever the viscosity drops at the lower branch it will have
a tendency to remain low. This result 
%is similar 
%to the conclusions
%obtained by Gammie \& Menou (1998) and 
supports the view that the angular
momentum transport due to MHD turbulence cannot be efficient when
the disk is in the quiescence and the assumption that $\alpha_{\rm
cold} \ll \alpha_{\rm hot}$ is justified.

\begin{figure}
\epsfxsize = 80 mm 
\epsfbox[18 144 610 700]{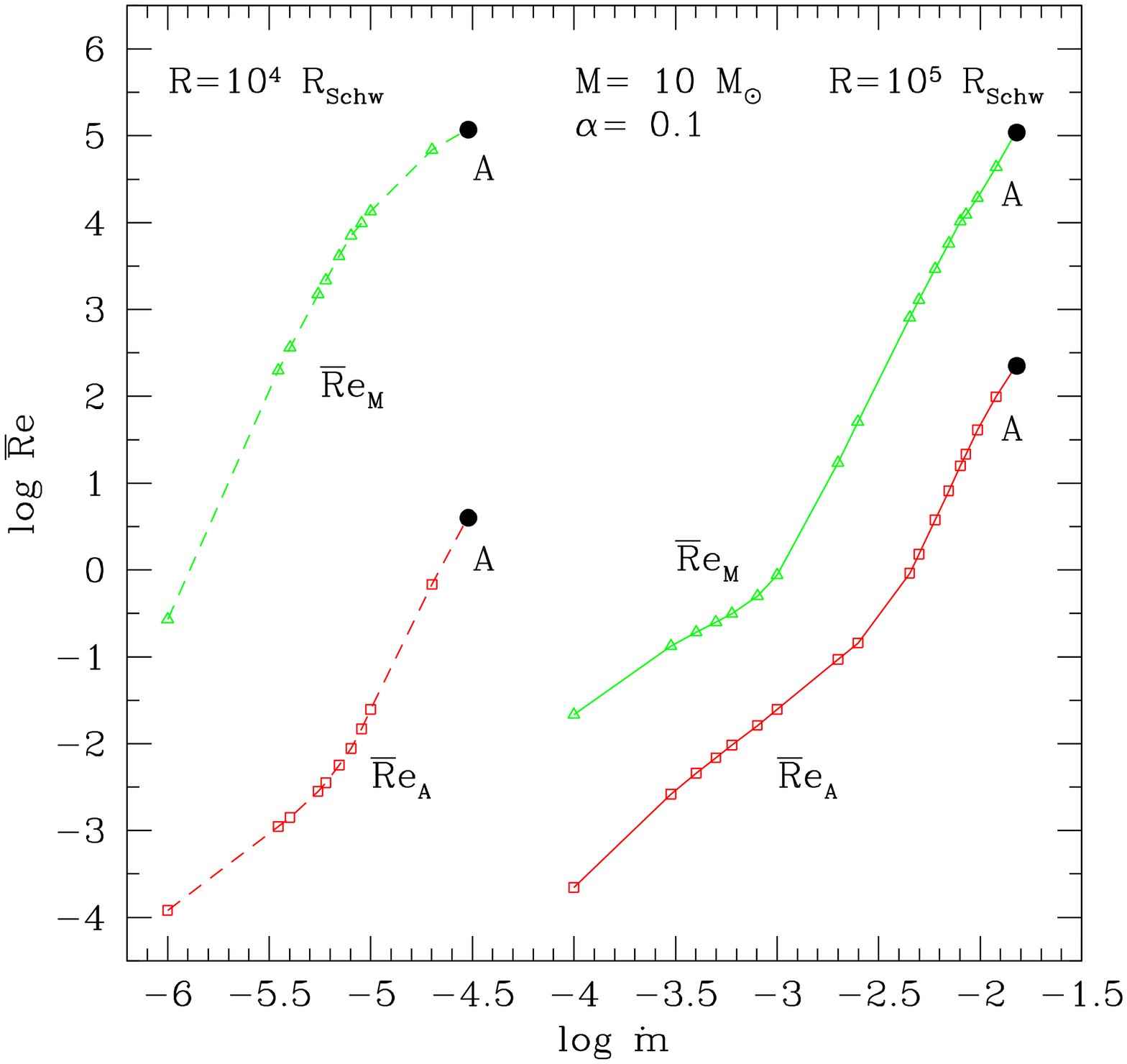}
%\plotone{f4.eps}
\caption{The density weighted magnetic and ambipolar Reynolds numbers 
calculated for black hole mass $M=10 M_{\odot}$, plotted for two disk radii
 $R=10^{4} R_{\rm Schw}$  (dashed line) and $R=10^{5}
R_{\rm Schw}$ (solid line), as a function of
the accretion rate in the range corresponding to the cold branch of
the S-curve. The viscosity is $\alpha = 0.1$.
\label{fig:re_trans}}
\end{figure}

We obtain a similar result for black hole X-ray transient systems with
M$_{\rm BH}$ $\sim 10 M_{\odot}$ as shown in Figure
\ref{fig:re_trans}.  The calculated magnetic and ambipolar Reynolds
numbers fall generally below the critical values for the quiescent disk in
these systems, regarding also the fact that this time there is no
difference in the location of the point $A$ caused by the change in $\alpha$.

\subsubsection{AGN accretion disks}

Now we consider accretion disks in AGN, e.g. systems harboring a
supermassive black hole.  Note, that for an accretion disk around a
supermassive black hole the temperatures are lower than in the stellar
mass black hole case.  Therefore the ionization instability zone is
located much closer to the central black hole and in our calculations
we have to consider radii of order of a few hundreds $R_{\rm Schw}$
instead of hundreds of thousands.

\begin{figure}
\epsfxsize = 80 mm 
\epsfbox[40 150 600 700]{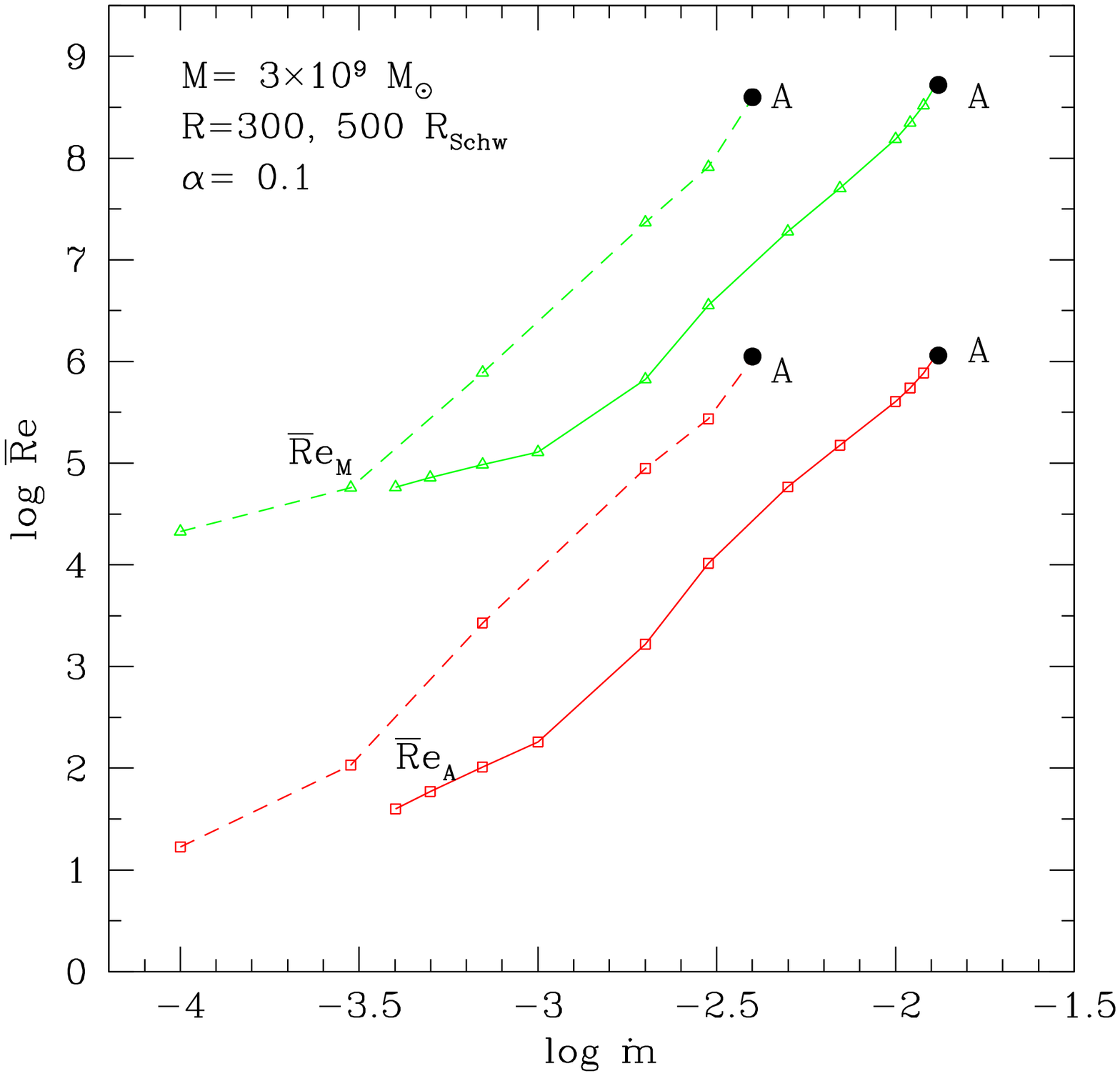}
%\plotone{f5.eps}
\caption{The density weighted magnetic and ambipolar Reynolds numbers 
calculated for black hole mass $M=3 \times  10^{9} M_{\odot}$,  
plotted for two disk radii $R=300 R_{\rm Schw}$ (dashed line) 
and $R=500 R_{\rm Schw}$ (solid line), as a function of the accretion rate
in the range corresponding to the cold branch of the S-curve 
(cf. Fig. \ref{fig:scurve}). The viscosity is $\alpha = 0.1$.
\label{fig:S_rey}}
\end{figure}

We calculate the Reynolds numbers for a range of
masses between $10^{6} M_{\odot}$ and $3 \times 10^{9}M_{\odot}$.  We
found that the Reynolds numbers are  orders of magnitude
above the critical values, for the most part of the lower branch, and
the greater the mass of a black hole, the larger $Re$.  In
Figure \ref{fig:S_rey} we plot magnetic and ambipolar Reynolds numbers
as a function of the accretion rate corresponding to the lower, cold
branch of the S-curve for
$3 \times 10^{9} M_{\odot}$.

Both Reynolds numbers clearly exceed the critical values: 10$^2$
and 10$^4$ for $Re_{\rm A}$ and $Re_{\rm M}$ respectively.  Therefore
only for very low accretion rates, $\dot m < 10^{-4}$,  MHD
turbulence in the disk would not develop and practically 
the entire cold branch
of the S-curve is MHD turbulent. This result only weakly depends on
the adopted viscosity parameter value. For $\alpha = 0.02$ both
Reynolds numbers had similar values to these presented in Figure
\ref{fig:S_rey} and were always greater than critical ($\bar Re_{\rm M} > 10^{6}$ 
and
$\bar Re_{\rm A} > 10^{5}$ at the disk radius of $\log R = 16.5$ [cm]) . 

\begin{figure}
\epsfxsize = 80 mm 
\epsfbox[40 150 600 680]{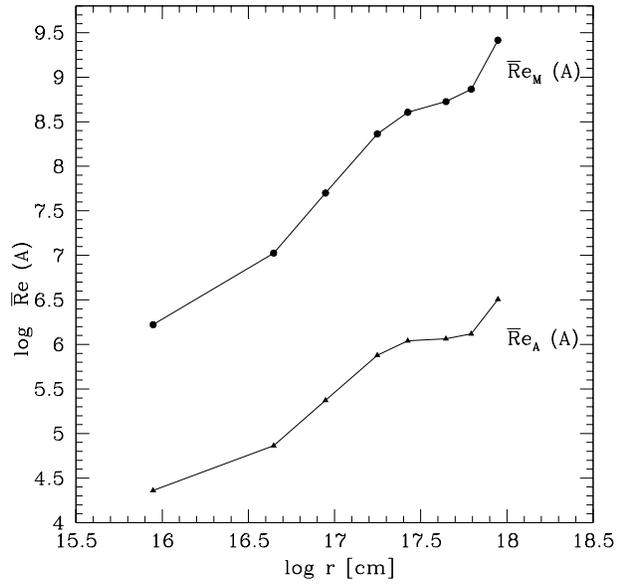}
%\plotone{f6.eps}
\caption{The radial profiles of the 
magnetic (circles) and ambipolar (triangles) Reynolds numbers in the
turning point $A$ on the S-curve. The points mark the values obtained
for 10, 50, 100, 200, 300, 500, 700 and 1000 $R_{\rm Schw}$ at each curve
respectively. The black hole mass is $M=3 \times 10^{9} M_{\odot}$ and
viscosity is $\alpha = 0.1$.
\label{fig:R_rey}}
\end{figure}

In order to understand whether the above result is valid for the
entire disk we calculate the Reynolds numbers at different locations
in the following way. 
At each radius we compute the entire S-curve, such as that in Figure
\ref{fig:scurve}, determine the position of the turning point $A$ and
calculate the Reynolds numbers $\bar Re_{\rm A}(A)$ and
$\bar Re_{\rm M}(A)$ at this point. Their values measured along the
lower branch of the S-curve are lower (equal) to $\bar Re_{\rm A}(A)$
and $\bar Re_{\rm M}(A)$.  We summarize this result in Figure
\ref{fig:R_rey}, plotting the Reynolds numbers against the radius.
Both $\bar Re_{\rm A}$ and $\bar Re_{\rm M}$ increase with radius
implying that MRI is stronger further out in the disk.  They decrease
towards smaller radii but do not drop below the critical values even
at 10 $R_{\rm Schw}$. 

%They remain high even for assumed lower viscosity
%parameter (e.g. for $\alpha=0.02$ we obtain $\bar Re_{\rm M} > 10^{6}$ and
%$\bar Re_{\rm A} > 10^{5}$ at the disk radius of $\log R = 16.5$ [cm]).

Our results support the view presented in Menou \& Quataert (2001) and
imply that in the thermally unstable AGN disks the viscosity should
not change between the cold and hot states of the disk
(i.e. $\alpha_{\rm cold} = \alpha_{\rm hot}$).

\section{Evaporation of the disk}
\label{sec:evap}

We concluded above that the viscosity parameter in AGN accretion disks
does not vary between the cold and hot states of the disk. As a result
the ionization instability will cause only small amplitude luminosity
fluctuations not really detectable in the observations (Siemiginowska,
Czerny \& Kostyunin 1996; Menou \& Quataert 2001). However, variations
in the local accretion rate induced by the instability may result in
the disk evaporation and affect the disk luminosity (Siemiginowska
1998). Here, we investigate the effects of the disk evaporation on the
overall evolution of thermally unstable disk. We assume that
$\alpha_{\rm cold} = \alpha_{\rm hot} = 0.1$.

\subsection{Evaporation radius}

The accretion flow in low luminosity systems can occur either via a
standard cold disk or via an advection dominated flow (ADAF). In the latter
the bulk of gravitational energy is advected into the central object
instead of being radiated (Ichimaru 1977; Narayan \& Yi 1994).  The
ADAF solution forms below a certain transition radius that depends on
the accretion rate. In classical ADAF model (Abramowicz et al. 1995;
Honma 1996; Kato \& Nakamura 1998) this relation is:
\begin{equation}
R_{\rm evap} = 1.9 \dot m^{-2} \alpha_{0.1}^{4} R_{\rm Schw}.
\label{eq:clasadaf}
\end{equation} 
The transition radius between the outer disk and inner ADAF 
was discussed in several papers (e.g. Liu et al. 1999; 
R\'o\.za\'nska \& Czerny 2000;
Menou et al. 2000; Liu et al. 2002; Meyer \& Meyer-Hofmeister 2002), under
various assumptions. Here we adopt the canonical approach.

The ionization instability can develop only when the standard thin
disk reaches the temperatures sufficient for the hydrogen ionization.
If accretion rates are low the disk can evaporate without being
ionized (e.g. without reaching point $A$ on the stability S-curve) and
the instability does not develop.
Therefore the evaporation radius cannot be very large for the
ionization to occur:
\begin{equation}
R_{\rm evap} < R_{\rm instab}.
\end{equation}

\begin{figure}
\epsfxsize = 80 mm 
\epsfbox[40 150 600 700]{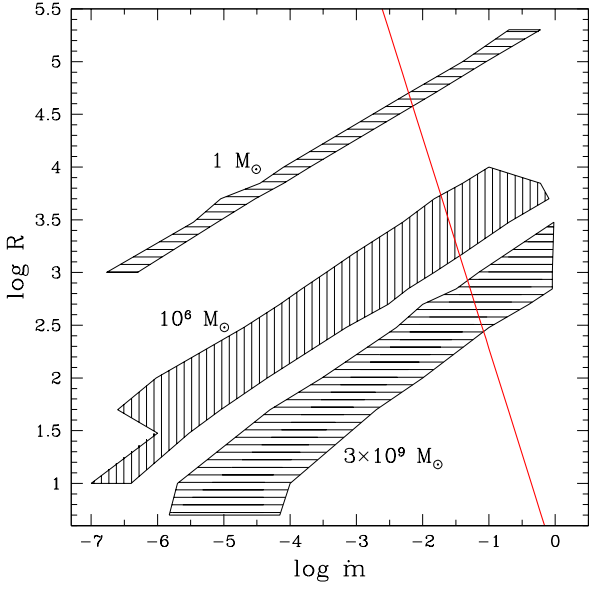}
%\plotone{f10.eps}
\caption{The radial extension of the ionization instability zone as a 
function of the accretion rate.  The contours correspond to
the turning points A and B on the S-curve (see Figure
\ref{fig:scurve}).  The thick solid line marks the transition radius,
resulting from the ADAF prescription (see text). The black hole mass is
 $M=1 M_{\odot}$ (top contour), $M=10^{6}
M_{\odot}$ (middle contour) and $M=3 \times 10^{9} M_{\odot}$ (bottom contour). 
The viscosity parameter is $\alpha = 0.1$.
\label{fig:topo}}
\end{figure}

In Figure \ref{fig:topo} we showed the radial extension of the unstable
ionization zone as a function of the accretion rate. The instability
cycle operates between the points $A$ and $B$ of the S-curve (cf. Figure
\ref{fig:scurve}), so we consider this entire range of accretion rates
to be unstable. 
In Figure \ref{fig:topo} we also indicate a location of $R_{\rm
evap}$ for the ADAF prescription 
(note, that the evaporation radius does not depend on mass of a black hole). 
If the disk
evaporates, only the unstable region {\it above} the  
solid line may exist and contribute to the total disk
luminosity.
The extension of the unstable region does not depend on $\alpha$ very
much, while the formula defining the evaporation radius strongly
depends on it.  Therefore, we should emphasize that our result may be
very sensitive to the adopted value of the viscosity
parameter. Possibly, depending on $\alpha$, the entire unstable region of the disk
evaporates, or, on the other hand, the evaporation radius is very
small and does not affect the instability. 

Because the location of the evaporation radius depends on
viscosity, if it can be determined observationally it may provide a
new method of estimating the value of the $\alpha$ parameter in
accreting objects.  For example, recently Nayakshin \& Sunyaev
(2003) proposed a new model explaining the X-ray flares in the
Galactic Center by the star-disk interactions. This model would favor
the case of the disk extending quite close to the central black
hole. However, the presence of the disk in the Galactic Center is
still unclear (see Quataert 2003 for a review).

\subsection{Luminosity of the disk with inner ADAF}

Below we study the evolution of the disk affected by the ionization
instability assuming that the viscosity parameter is constant, 
as favored for accreting supermassive black holes. We used the method
described in Siemiginowska, Czerny \& Kostynuin (1996) and consider
the two cases: (i) the accretion disk extends down to the marginally
stable orbit and (ii) the central part of the disk is evaporated and
forms an ADAF, while the standard thick disk remains at the
outer radii.  Since the accretion efficiency of an ADAF is low, 
we assume that this part of the flow does not contribute to the luminosity.

\begin{figure}
\epsfxsize = 80 mm 
\epsfbox[40 150 600 700]{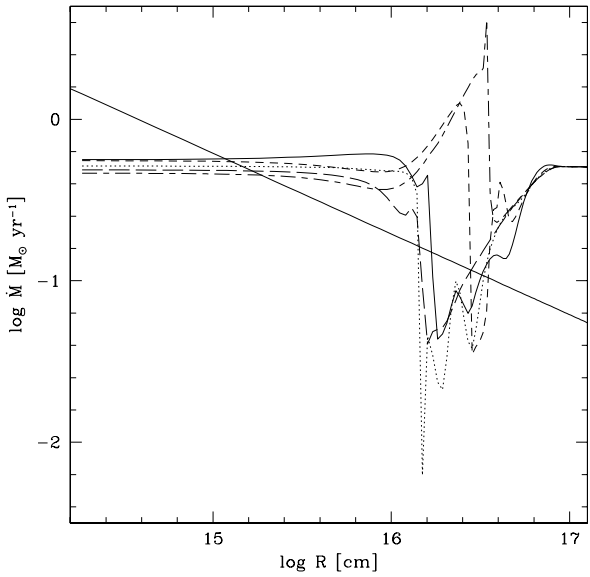}
%\plotone{f11.eps}
\caption{The local accretion rate during the cycle of
the disk evolution caused by the ionization instability.
The thick solid line 
marks the critical accretion rate, below which the disk is evaporated, 
resulting from the ADAF prescription (see Eq. \ref{eq:clasadaf}). 
The black hole mass is $M=10^{8} M_{\odot}$ and viscosity $\alpha = 0.1$.
\label{fig:mcrit}}
\end{figure}

In order to determine the disk evaporation radius during the evolution
we compare the local accretion rate with the critical (evaporation)
accretion rate. We use the classical ADAF prescription, expressed by
 Equation \ref{eq:clasadaf}. In Figure \ref{fig:mcrit} we show
radial profiles of the local accretion rate in several snapshots
during one cycle of the disk instability. The straight solid line
marks the critical accretion rate. Whenever the local accretion rate
in the disk drops below this critical value, the disk evaporates and
 ADAF is formed.  Hereafter, in our time dependent calculations we
assume that the ADAF transition radius is the outermost radius at
which the disk evaporates. Therefore even though the local accretion
rate may be equal to the critical value at several locations in the
disk, only the outermost one determines the transition radius and the
standard disk cannot rebuilt itself below this radius.

\begin{figure}
\epsfxsize = 80 mm 
\epsfbox[40 150 560 700]{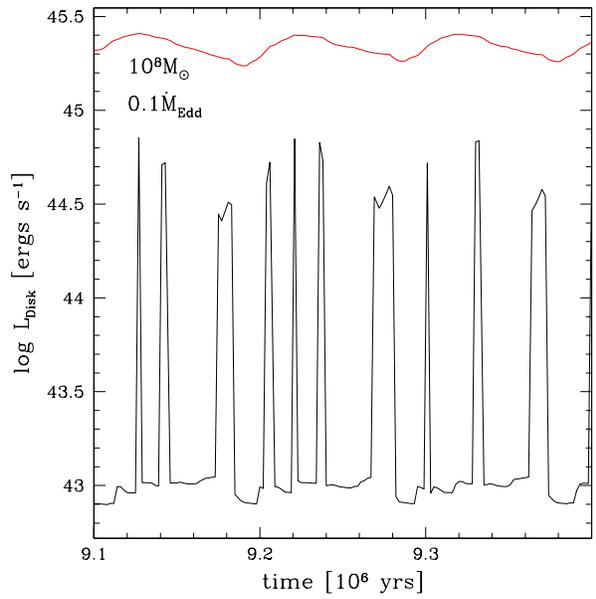}
%\plotone{f12.eps}
\caption{The disk 
lightcurve during the evolution caused by the ionization instability
with constant viscosity assumed ($\alpha = 0.1$).  The upper curve
shows the luminosity flickering resulting from the evolution of the
whole disk. The lower curve shows the sharp outbursts of the disk,
which inner part is evaporated.  The black hole mass is $M=10^{8}
M_{\odot}$ and the global accretion rate is 0.1 $\dot M_{\rm Edd}$.
\label{fig:ldisk_adaf}}
\end{figure}

In Figure~\ref{fig:ldisk_adaf} we present lightcurves for the two
models: (i) the standard disk and (ii) the disk with inner ADAF.  When
the disk extends all the way to the marginally stable orbit, as
considered in the previous time-dependent calculations, the luminosity
variations are small and only a small flickering ($\Delta
\log L_{\rm disk} \sim 0.2$) can be seen in the resulting lightcurve.
This is indicated by the upper solid line in Figure \ref{fig:ldisk_adaf}.

However, the situation changes dramatically, when the evaporation of
the inner disk is taken into account. In this case the ionization
instability leads to strong, large-amplitude outbursts.  These
outbursts are due to the large variations of the inner radius of the
accretion disk (e.g. the size of the ADAF), as determined by the
critical value of the local accretion rate.  The outburst peak
corresponds to the moment, when the inner disk radius is small and the
entire accretion disk contributes to the total luminosity. The
quiescent low luminosity state, on the other hand, is achieved when
the local accretion rate drops below the critical value at the large
distance from the center and most of the inner disk is evaporated.

\section{Discussion and conclusions}
\label{sec:diss}

We studied the effects of MHD turbulence and the efficiency of
angular momentum transport in the quiescent accretion disks.  We
analyzed the full disk vertical structure, which enabled us to
determine accurately the coupling of the accreting gas to the magnetic
field over the entire disk height.

Confirming the previous results of Menou \& Quataert (2001), we find
that only in the case of accretion disks in binary systems,
e.g. around white dwarfs (CVs) or stellar mass black holes (SXTs), the
timescale of the magnetic field decay is too short for the field to be
frozen into the gas. Therefore, MHD turbulence does not operate
efficiently and the large amplitude outbursts observed in these
sources can be accounted for by the thermal ionization instability
cycle in which $\alpha_{\rm cold} \ll \alpha_{\rm hot}$.  On the
other hand, in the case of accreting supermassive black holes in AGN
the gas in the disk is well coupled to the magnetic field. 

This effect
($Re_{\rm M} \gg 1$ and $Re_{\rm A} \gg 1$) can be only slightly reduced
by encounters of free electrons with charged dust particles.
In principle, the presence of dust in the disk can affect the number density 
of free electrons. However, the grains evaporate if the temperature exceeds
$\sim 1800-2000$ K, which is the case near the disk equatorial plane. 
As a result, the role of the dust in suppressing MHD turbulence is limited
only to the uppermost disk layers and cannot have a global effect.
We conclude, that the AGN accretion disks in quiescence have
comparable angular momentum transport efficiency to the disks in
outburst.

Therefore, we studied the ionization instability assuming a constant value
of the viscosity parameter $\alpha_{\rm cold} = \alpha_{\rm hot}$.
We took into account the evaporation of the innermost parts of the disk
during the quiescence in AGN accretion disks.
The transition to an ADAF below the evaporation
radius during the evolution of the instability
results in large luminosity variations ($\Delta \log L_{\rm disk} \sim
2$). Because only
the outer parts of the disk ($R > R_{\rm evap} \sim 10^{3} R_{\rm
Schw}$) contribute to the total disk emission the minimum luminosity 
obtained within the instability cycle is lower than in the standard 
disk with no evaporation.
We note, that also
during the outburst the evaporation is important and
ADAF remains in the innermost regions of the disk ($R_{\rm evap} \sim
25 R_{\rm Schw}$), even at quite a high global accretion rate, $\dot M
= 0.1 \dot M_{\rm Edd}$.  The evaporation radius, however, strongly
depends on the adopted value of the viscosity parameter $\alpha$
(cf. Equation \ref{eq:clasadaf}).

The $\alpha$-disks in AGN are self-gravitating beyond a certain
radius (Paczy\'nski 1978; Hure 1998).  A typical distance of the
self-gravitating zone is about a thousand Schwarzschild radii (Hure
2000). However, the greater the black hole mass and lower the
parameter $\alpha$, the closer to the center this zone is located. 
We can include a self-gravitation term 
equal to $-4\pi G \Sigma$ in Equation \ref{eq:hydro} of the disk
vertical structure. This additional term 
modifies the S-curve, so there is a single 'wiggle' on the middle
branch of the curve, instead of two. 
The middle branch is also broader
and reaches lower surface densities. 
The instability happens for a narrower range of effective temperatures,
while the more strongly curved unstable middle branch covers broader 
range  of surface density values. Therefore, the amplitude of the
outburst is not expected to be significantly modified in our model
by the presence of self-gravity.
However, whether the model actually can be extended into 
the self-gravity region, remains an open question. In particular,
our description of convection, which influences
the shape of the S-curve, may not be applicable in case of a 
self-gravitating disk. Also, MRI was not extensively studied in the
self gravity region (but see Fromang et al. 2003) 
and the use of viscosity parameter $\alpha$
may not be appropriate here. Instead, self-gravity may either provide 
a mechanism of angular momentum transfer (e.g. Paczy\'nski 1978),
or lead to the disk fragmentation and star formation (Collin \&
Zahn 1999). The physics of those processes is poorly known. Therefore the use 
of simple models like $\alpha$-disks and comparison of their predictions
with observed behavior of AGN seems to be justified.

The variability amplitudes predicted by our model are about two orders
of magnitude and support the observational evidence of intermittent
activity of AGN.  This kind of activity has been proposed to explain
the formation of giant radio galaxies (Subrahmanyan et al. 1996) and
the extended radio structures in the Giga-Hertz Peaked Spectrum (GPS)
radio sources (Baum et al. 1990; Schoenmakers et al. 1998). The timescales
of episodic activity in quasars were recently estimated by Martini \& Schneider (2003).
Recent
observations of Compact Steep Spectrum (CSS) sources indicate that the
evolutionary path of radio loud AGN is consistent with an on-off
activity with duration governed by the mass of the central black hole
(Marecki et al. 2003).  Also the black hole masses implied by the HST
observations of normal galaxies and AGN suggest that every galaxy must
undergo an active phase in its life (Ferrarese et al. 2001).

The estimated timescales of the activity cycle are within $10^{5}$ to
10$^{7}$~years (e.g. Reynolds \& Begelman 1997 for GPS sources). Di
Matteo et al. (2003) proposed a feedback mechanism to explain the
cyclic activity in M87, which occurs on a timescale of 10$^{7}$ years.
In our time-dependent calculations, we adopted a simplified approach
to the disk evaporation process, and therefore underestimated the
repetition timescale.  This is because we do not consider the time of
the disk reconstruction after it has been evaporated, which increases
the viscous timescale by a factor of $\sim 2$.
Assuming that the
outburst starts at the outer edge of the instability zone we obtain
the viscous timescale of $t_{\rm visc} = 1/(\alpha\Omega)\times (r/H)^{2}
\approx 0.9$ Myrs for $M_{\rm BH} = 10^{8} M_\odot$.
Thus the outbursts will be less frequent than these presented in
Figure \ref{fig:ldisk_adaf}.
However, because during the outburst there is no ADAF and 
the entire disk is present,
the decay timescale is calculated properly and the
 duration of an outburst is correct.
On the other hand, the outburst profiles will be less steep than the ones
calculated here, since the disk density rises gradually at the
beginning of the instability cycle. Although these two effects are
important, the extension of the unstable region is not very large in
comparison with the disk size and the uncertainty involved in our
calculations is not crucial. The detailed model of the time evolution
of the accretion disk with transition to ADAF is clearly worth
investigating and we address it to the future work.

{\sl Acknowledgments} AJ is grateful to F. Meyer and
E. Meyer-Hofmeister for interesting discussion. This work was
supported in part by grants 2P03D00322 and 2P03D00124 of the Polish
State Committee for Scientific Research.
Partial support for this work was provided by the National Aeronautics
and Space Administration through Chandra Award Number GO-01164X issued
by the Chandra X-Ray Observatory Center, which is operated by the
Smithsonian Astrophysical Observatory for and on behalf of NASA under
contract NAS8-39073.

\section*{References}
%\begin{thebibliography}{}
%\bibitem[]{A2} 

\Referencja{Abramowicz M.A., Chen X., Kato S., Lasota J.-P., Regev O.,
1995,   ApJ, {\bf 438}, L37}
%\bibitem[]{A1} 
\Referencja{Alexander D.R., Johnson H.R., Rypma R.L., 1983, ApJ, {\bf 272}, 773}

%\bibitem[]{B1} 
\Referencja{Balbus S.A., Hawley J.F., 1991, ApJ, {\bf 376}, 214}
%\bibitem[]{B2} 
\Referencja{Balbus S.A., Hawley J.F., 1998, Accretion Processes in Astrophysical Systems: Some Like it Hot!, eds. Holt S.S. \& Kallman T.R., AIP Conf. Proceedings, {\bf 431}, p. 79
}
%\bibitem[]{B3} 
\Referencja{Baum S.A., O'Dea C.P., de Bruyn A.G., Murphy D.W., 1990, A\&A, {\bf 232}, 19}
%\bibitem[]{B4} 
\Referencja{Begelman M.C., 1985, Astrophysics of Active Galaxies and Quasi-Stellar Objects, p. 411 (Mill Valley: University Science Books) }
%\bibitem[]{B5} Bowers R.L., Deeming T., 1984, Astrophysics. Vol. 2: Interstellar matter and galaxies,  (Boston:Jones \& Bartlett)
%\bibitem[]{C1} 
\Referencja{Cannizzo J.K., Ghosh P.,  Wheeler J.C., 1982, ApJ, {\bf 260}, L83}
%\bibitem[]{C2} 
\Referencja{Cannizzo J.K., 1993, ApJ, {\bf 419}, 318}
%\bibitem[]{C3} 
\Referencja{Clarke C.J., 1988, MNRAS, {\bf 235}, 881}
%\bibitem[]{C4} 
\Referencja{Collin S., Zahn J.-P., 1999, A\&A, {\bf 344}, 433}
%\bibitem[]{D1} 
\Referencja{Di Matteo T., Allen S.W., Fabian A.C., Wilson A.S., Young A.J., 2003,
ApJ, {\bf 582}, 133}
%\bibitem[]{D2} 
\Referencja{Dubus G., Hameury J.-M., Lasota J.-P., 2001, A\&A, {\bf 373}, 251}
%\bibitem[]{F1} 
\Referencja{Ferrarese L., Pogge R.W., Peterson B.M., Merritt D., Wandel A., Joseph C.L., 2001, ApJ, {\bf 555}, L79}
%\bibitem[]{F2} 
\Referencja{Fromang S., de Villiers J.P., Balbus S.A., 2003 (astro-ph/0309816)}
%\bibitem[]{G1} 
\Referencja{Gammie C.F., Menou K., 1998, ApJ, {\bf 492}, L75}
%\bibitem[]{H1} 
\Referencja{Hatziminaoglou E., Siemiginowska A., Elvis M., 2001, ApJ, {\bf 547}, 90}
%\bibitem[]{H2} 
\Referencja{Hawley J.F., Gammie C.F., Balbus S.A., 1996, ApJ, {\bf 464}, 690}
%\bibitem[]{H3} 
\Referencja{Hawley J.F., Stone J.M., 1998, ApJ, {\bf 501}, 758}
%\bibitem[]{H4} 
\Referencja{Honma F., 1996, PASJ, {\bf 48}, 77}
%\bibitem[]{H5} 
\Referencja{Hure J.-M., 1998, A\&A, {\bf 337}, 625}
%\bibitem[]{H6} 
\Referencja{Hure J.-M., 2000, A\&A, {\bf 358}, 378}
%\bibitem[]{I1} 
\Referencja{Ichimaru S., 1977, ApJ, {\bf 214}, 840
}
%%\bibitem[]{J1} Jackson J.D., 1975, Classical Electrodynamics, (New York:Wiley)
%\bibitem[]{J2} 
\Referencja{Janiuk, A., Czerny, B., \& Siemiginowska, A., 2002, ApJ, {\bf 576}, 908} 
%%\bibitem[]{K1} Karovska M., Fabbiano G., Nicastro F., Elvis M.,  
%Kraft R.P., Murray S.S., 2002, ApJ, {\bf 577}, 114
%\bibitem[]{K2} 
\Referencja{Kato S., Nakamura K.E., 1998, PASJ, {\bf 50}, 559}
%\bibitem[]{L1} 
\Referencja{Lasota J.-P., 2001, NewAR, {\bf 45}, L449}
%\bibitem[]{L2} 
\Referencja{Lin D.N.C., Shields G.A., 1986, ApJ, {\bf 305}, 28}
%\bibitem[]{L3} 
\Referencja{Liu B.F., Yuan W., Meyer F., Meyer-Hofmeister E., Xie G.Z.,
1999, ApJ, {\bf 527}, L17}
%\bibitem[]{L4} 
\Referencja{Liu B.F., Mineshige S., Meyer F., Meyer-Hofmeister E.,
Kawaguchi T., 2002, ApJ, {\bf 575}, 117}
%\bibitem[]{M0} 
\Referencja{Marecki A., Spencer R.E., Kunert M., 2003, PASA, {\bf 20}, 46}
%\bibitem[]{M8} 
\Referencja{Martini P., Schneider P., 2003 (astro-ph/0309650)}
%\bibitem[]{M1} 
\Referencja{Menou K., Hameury J.-M., Lasota J.-P., Narayan R., 2000, MNRAS, {\bf 314}, 498}  
%\bibitem[]{M2} 
\Referencja{Menou K., Quataert E., 2001, ApJ, {\bf 552}, 204}
%\bibitem[]{M3} 
\Referencja{Meyer F., Meyer-Hofmeister E., 1981, A\&A, {\bf 104}, L10}
%\bibitem[]{M4} 
\Referencja{Meyer F., Meyer-Hofmeister E., 1982, A\&A, {\bf 106}, 34}
%\bibitem[]{M5} 
\Referencja{Meyer F., Meyer-Hofmeister E., 1994, A\&A, {\bf 288}, 175}
%\bibitem[]{M6} 
\Referencja{Meyer F., Meyer-Hofmeister E., 2002, A\&A, {\bf 392}, L5}
%\bibitem[]{M7} 
\Referencja{Mineshige S., Shields G.A., 1990, ApJ, {\bf 351}, 47}
%\bibitem[]{N1} 
\Referencja{Narayan R., Yi I., 1994, ApJ, {\bf 428}, L13}
%\bibitem[]{N2} 
\Referencja{Nayakshin S., Sunyaev R., 2003, MNRAS, {\bf 343}, L15}
%\bibitem[]{P0} 
\Referencja{Parker E.N., 1979, Cosmical Magnetic Fields, (Oxford:Clarendon Press)}
%\bibitem[]{P1}
\Referencja{ Paczy\'nski B., 1978, Acta Astron., {\bf 28}, 91}
%\bibitem[]{P2}
\Referencja{ Pojma\' nski G., 1986, Acta Astron. {\bf 36}, 69}
%\bibitem[]{Q1} 
\Referencja{Quataert, 2003 in Astron. Nachr., Vol. {\bf 324}, No. S1  
(astro-ph/0304099) }
%\bibitem[]{R1} 
\Referencja{Rees M.J., Begelman M.C., Blandford R.D., Phinney E.S., 1982, Nature, {\bf 295}, 17}
%\bibitem[]{R2} 
\Referencja{Reynolds C.S., Begelman M.C., 1997, ApJ, {\bf 487}, L135}
%\bibitem[]{R3} 
\Referencja{R\' o\. za\' nska A., Czerny B., \. Zycki P.T., Pojma\' nski G.,
 1999, MNRAS, {\bf 305}, 481}
%\bibitem[]{R4} 
\Referencja{R\' o\. za\' nska A., Czerny B., 2000, A\&A, {\bf 360}, 1170}
%\bibitem[]{S1} 
\Referencja{Schoenmakers A.P., de Bruyn A.G., Rottgering H.J.A., van der Laan H., 1999, A\&A, {\bf 341}, 44}
%\bibitem[]{S2} 
\Referencja{Seaton M.J., Yan Y., Mihalas D., Pradhan A.K., 1994, MNRAS, {\bf 266}, 805}
%\bibitem[]{S3} 
\Referencja{Shakura N.I., Sunyaev R.A., 1973, A.\&A., {\bf 24}, 337}
%\bibitem[]{S4} 
\Referencja{Siemiginowska A., Czerny B., Kostyunin V., 1996, ApJ, {\bf 458}, 491}
%\bibitem[]{S5} 
\Referencja{Siemiginowska, A.\ 1998, Accretion Processes in Astrophysical Systems: Some Like it Hot!, eds.  Holt S.S. \& Kallman T.R., AIP Conference Proceedings {\bf 431}, p. 211 }
%\bibitem[]{S6} 
\Referencja{Smak J. 1982, Acta Astron. {\bf 32}, 199}
%\bibitem[]{S7} 
\Referencja{Smak J. 1984, Acta Astron. {\bf 34}, 161}
%%\bibitem[]{S8} Spitzer L. Jr., 1941, ApJ, {\bf 93}, 369
%\bibitem[]{S9} 
\Referencja{Subrahmanyan R., Saripalli L., Hunstead R.W., 1996, MNRAS, {\bf 279}, 257}

%\end{thebibliography}
\clearpage

\clearpage

\end{document}